\RequirePackage{fix-cm}
\documentclass[twocolumn]{svjour3}          
\smartqed  
\usepackage{graphicx}
\usepackage{amsfonts}
\usepackage{amsmath}
\usepackage{amssymb}
\usepackage[usenames]{color}
\journalname{Journal of Superconductivity and Novel Magnetism}
\begin{document}

\title{Multiband superconductivity due to the electron - LO-phonon interaction in
strontium titanate and on a SrTiO$_{3}$/LaAlO$_{3}$ interface}

\titlerunning{Multiband superconductivity in strontium titanate}        

\author{S. N. Klimin  \and
        J. Tempere \and
        J. T. Devreese \and
        D. van der Marel
}

\authorrunning{S. N. Klimin, J. Tempere, J. T. Devreese and D. van der Marel} 

\institute{S. N. Klimin
    \and
    J. Tempere
    \and
    J. T. Devreese \at
    Theorie van Kwantumsystemen en Complexe Systemen (TQC),
    Universiteit Antwerpen, Universiteitsplein 1, BE-2610 Antwerpen, Belgium\\
    \email{sergei.klimin@uantwerpen.be}
    \and
    J. Tempere \at
    Lyman Laboratory of Physics, Harvard University, Cambridge,
    Massachusetts 02138, USA
    \and
    D. van der Marel \at
    D\'{e}partement de Physique de la Mati\`{e}re Condens\'{e}e, Universit\'{e} de
    Gen\`{e}ve, CH-1211 Gen\`{e}ve 4, Switzerland
}

\date{Received: date / Accepted: date}

\maketitle

\begin{abstract}
In strontium titanate, the Fr\"{o}hlich electron -- LO-phonon interaction
dominates the electron response and can also provide superconductivity.
Because of high LO-phonon frequencies in SrTiO$_{3}$, the superconducting
system is non-adiabatic. We demonstrate that the dielectric function approach
is an adequate theoretical method for superconductivity in SrTiO$_{3}$ and on
the SrTiO$_{3}$-LaAlO$_{3}$ interface. The critical temperatures are
calculated using realistic material parameters. The obtained critical
temperatures are in line with experimental data both for bulk and interface
superconductivity. The present method explains the observed multi-dome shape
of the critical temperature in SrTiO$_{3}$ as a function of the electron
concentration due to multiband superconductivity.
\end{abstract}

\section{Introduction}

The discovery of a highly conducting two-dimensional electron gas at the
interface between two insulators, SrTiO$_{3}$ and LaAlO$_{3}$ \cite{Ohtomo}
stimulated intense experimental and theoretical studies which led to the
observation of fascinating phenomena such as superconductivity
\cite{Reyren,Caviglia,Richter,Boschker2015,Boschker2016} coexisting with
ferromagnetism \cite{Li,Bert,Dikin} and spin-orbit coupling
\cite{Caviglia2010,Zhong}. These effects are related to unique properties of
strontium titanate. The superconducting phase transition in strontium titanate
and on SrTiO$_{3}$-based interfaces occurs at low temperatures combined with
low carrier densities, so that it was called \textquotedblleft the most dilute
superconductor\textquotedblright\ \cite{Lin}. Recent experimental studies
demonstrated the possibility to control parameters of the electron gas at the
LaAlO$_{3}$/SrTiO$_{3}$ interface by an external electric field
\cite{Caviglia,Richter,Boschker2015,Boschker2016}. In particular, this allow
to study the dependence of the superconducting transition temperature on the
electron concentration. Consequently, these investigations have reopened the
discussion about the mechanism of superconductivity in bulk strontium titanate.

SrTiO$_{3}$ is a strongly polar crystal, and the Fr\"{o}hlich electron --
LO-phonon interaction dominates in the electron response of SrTiO$_{3}$
\cite{KDM2010}. Recent theoretical studies \cite{Ruhman,Edge} argued that
other mechanisms than the electron-LO phonon coupling are responsible for
superconductivity in SrTiO$_{3}$. Nevertheless, the phonon-mediated
electron-electron attraction is still considered as the most likely candidate
to provide superconductivity in strontium titanate
\cite{Koonce,Takada1,DKTM2014,Gorkov,Gorkov2}.

Because the LO-phonon frequencies are high with respect to both the thermal
energy and the Fermi energy, the superconducting system is strongly
non-adiabatic. The adequate theoretical method for superconductivity in the
non-adiabatic electron -- LO-phonon system is the dielectric function approach
\cite{Takada1,Kirzhnits,Takada2}. We apply the dielectric function approach as
a unique method to treat superconductivity both at the SrTiO$_{3}$-LaAlO$_{3}$
interface \cite{DKTM2014} and in bulk doped strontium titanate. This study is
performed using recent results for the band structure of SrTiO$_{3}$
\cite{VDM2008,VDM2011,Meevasana}. The critical temperatures are calculated
without fitting. For the calculations, we use well-established material
parameters, except for the acoustic deformation potential of strontium
titanate, for which the values reported in the literature show a considerable
spread \cite{Janotti,Morozovska}. The calculated critical temperatures are
compared with available experimental data.

\section{Superconductivity in complex oxides and their interfaces within the
dielectric function method}

In view of recent discussions, it is relevant to revisit the theoretical
treatment of superconductivity in bulk strontium titanate. There exist
different approaches to treat the superconductivity in polar crystals due to
the electron -- LO-phonon interaction, see, e. g., Refs.
\cite{Koonce,Takada1,DKTM2014,Gorkov,Kirzhnits}. In the present work, the
superconducting transition temperatures in $n$-doped SrTiO$_{3}$ are
calculated within the dielectric function approach
\cite{Kirzhnits,Takada1,Takada2}. We can recalculate the critical temperatures
in bulk strontium titanate using the dielectric function method accounting for
the most recent results \cite{VDM2008,VDM2011,Meevasana} on the band structure
and optical-phonon spectrum of strontium titanate.

Recently, an alternative approach for the description of superconductivity in
complex oxides and their interfaces has been developed in Refs.
\cite{Gorkov,Gorkov2}. It is stated in these works that the electron-electron
interaction mediated by the LO phonons is sufficient to overcome the Coulomb
repulsion in strontium titanate and can lead therefore to an effective
attraction potential. This question requires careful analysis. Below, we argue
that the approach of \cite{Gorkov,Gorkov2} in which the LO-phonon mediated
electron-electron interaction relates to the static dielectric constant is
insufficient to describe superconductivity in strontium titanate.

The matrix element for the effective electron-electron interaction provided by
phonons is modeled in Ref. \cite{Gorkov} by a two-particle potential in the
Bardeen -- Pines form \cite{BP}, which for a single-mode polar crystal is
given by:%
\begin{equation}
\Gamma\left(  \mathbf{p},E_{n}|\mathbf{k},E_{m}\right)  =\frac{4\pi e^{2}%
}{\varepsilon_{\infty}q^{2}}-\frac{2\omega_{L}\left\vert V_{\mathbf{q}%
}\right\vert ^{2}}{\omega_{L}^{2}+\left(  E_{n}-E_{m}\right)  ^{2}}.
\label{mel2}%
\end{equation}
where $\mathbf{q}=\mathbf{p}-\mathbf{k}$ and $E_{n}-E_{m}$ are, respectively,
the momentum and frequency the electrons exchange upon scattering,
$\varepsilon_{\infty}$ and $\varepsilon_{0}$ are high-frequency and static
dielectric constants, $\omega_{L}$ is the LO-phonon frequency, and $\left\vert
V_{\mathbf{q}}\right\vert ^{2}=4\pi\alpha\left(  \omega_{L}/q\right)
^{2}\left(  2m_{b}\omega_{L}\right)  ^{-1/2}$ is the squared modulus of the
electron-phonon interaction amplitude with the band mass $m_{b}$ and the
electron-phonon coupling constant $\alpha=e^{2}\left(  m_{b}/2\omega
_{L}\right)  ^{1/2}\left(  1/\varepsilon_{\infty}-1/\varepsilon_{0}\right)  $.
In a \emph{multimode} polar crystal with $n$ optical-phonon branches, like
SrTiO$_{3}$, the matrix element (\ref{mel2}) is extended in a straightforward
way:%
\begin{equation}
\Gamma\left(  \mathbf{p},E_{n}|\mathbf{k},E_{m}\right)  =\frac{4\pi e^{2}%
}{\varepsilon_{\infty}q^{2}}-\sum_{j=1}^{n}\frac{2\omega_{L,j}\left\vert
V_{\mathbf{q},j}\right\vert ^{2}}{\omega_{L,j}^{2}+\left(  E_{n}-E_{m}\right)
^{2}}. \label{mel1}%
\end{equation}
where $V_{\mathbf{q},j}$ is the electron-phonon interaction amplitude for a
$j$-th LO-phonon branch. The partial coupling strengths for any phonon branch
in a continuum approach for the electron-phonon interaction can be determined
as described in Refs. \cite{T1972,KDM2010}. The amplitudes of the
electron-phonon interaction for a large (Fr\"{o}hlich) polaron in a multimode
crystal are:%
\begin{equation}
\left\vert V_{\mathbf{q}j}\right\vert ^{2}=\frac{4\pi e^{2}}{q^{2}}\left.
\left(  \frac{\partial\varepsilon\left(  \omega\right)  }{\partial\omega
}\right)  ^{-1}\right\vert _{\omega=\omega_{L,j}}, \label{rrr3}%
\end{equation}
where $\varepsilon\left(  \omega\right)  $ is a dielectric function for a crystal.

Let us substitute in (\ref{rrr3}) the frequently used model dielectric
function for a multimode polar crystal \cite{T1972,mmc3}:\textrm{ }%
\begin{equation}
\varepsilon\left(  \omega\right)  =\varepsilon_{\infty}\prod_{j=1}^{n}\left(
\frac{\omega^{2}-\omega_{L,j}^{2}}{\omega^{2}-\omega_{T,j}^{2}}\right)
,\label{DF}%
\end{equation}
whose zeros and poles correspond to the LO and TO phonon frequencies $\left\{
\omega_{L.j},\omega_{T.j}\right\}  $, respectively. This dielectric function
is the result of the extension of the Born-Huang approach \cite{BH1954} to
multimode polar crystals. A particular consequence of (\ref{DF}) is the
extension of the Lydanne-Sachs-Teller (LST) relation: $\varepsilon_{\infty
}/\varepsilon_{0}=\prod_{j=1}^{n}\left(  \omega_{T,j}^{2}/\omega_{L,j}%
^{2}\right)  $. One can explicitly check the exact analytic equality:%
\begin{equation}
1-\sum_{j=1}^{n}\left(  1-\frac{\omega_{T,j}^{2}}{\omega_{L,j}^{2}}\right)
\prod_{j^{\prime}\neq j}\left(  \frac{\omega_{L,j}^{2}-\omega_{T,j^{\prime}%
}^{2}}{\omega_{L,j}^{2}-\omega_{L,j^{\prime}}^{2}}\right)  =\frac
{\varepsilon_{\infty}}{\varepsilon_{0}}.\label{equ1}%
\end{equation}
In the antiadiabatic case, as assumed in Ref. \cite{Gorkov}, $\omega_{L,j}%
\gg\left\vert E_{n}-E_{m}\right\vert $ and hence one can omit $\left\vert
E_{n}-E_{m}\right\vert $ in the denominator of the phonon Green function in
(\ref{mel1}). Accounting for (\ref{equ1}), the matrix element of the effective
electron-electron interaction provided by the interplay of the Coulomb
repulsion and the LO-phonon-mediated attraction in the antiadiabatic limit is
reduced to the expression, \emph{which is the same for single-mode and
multimode crystals}:%
\begin{equation}
\left.  \Gamma\left(  \mathbf{p},E_{n}|\mathbf{k},E_{m}\right)  \right\vert
_{\omega_{L,j}\gg\left\vert E_{n}-E_{m}\right\vert }=\frac{4\pi e^{2}%
}{\varepsilon_{0}q^{2}}.\label{gam1}%
\end{equation}
Thus the effective attraction in a multimode polar crystal in the
antiadiabatic limit results in the replacement of $\varepsilon_{\infty}$ to
$\varepsilon_{0}$ where $\varepsilon_{0}$ accounts for the polarization due to
\emph{all} phonon modes. After this partial compensation of the Coulomb
repulsion, \emph{no more LO phonons remain to provide additional attraction}.
Moreover, beyond the antiadiabatic limit, when we do not neglect $\left(
E_{n}-E_{m}\right)  ^{2}$ in the denominator of (\ref{mel1}), the sum over
LO-phonon modes in (\ref{mel1}) becomes even smaller. Thus, as long as the
LO-phonon-mediated effective electron-electron interaction is modeled by a
\emph{static} two-particle potential, it is not sufficient to provide
superconductivity in a polar (both single-mode and multimode) crystal. The
physical reason for this conclusion is quite transparent. In the antiadiabatic
limit, the dielectric response of a crystal on the electron is the same as on
a static charge, resulting in the static dielectric function. The static
dielectric function in the long-wavelength limit must be positive from the
stability condition. Therefore the total effective interaction remains
repulsive. Note that the same conclusion on the work \cite{Gorkov} has been
recently independently made in Ref. \cite{Ruhman}.

Also in layered structures \cite{Gorkov2}, the phonon-mediated attraction in
the antiadiabatic limit results in static dielectric image forces which cannot
overcome the Coulomb repulsion. We can conclude that an effective
LO-phonon-mediated attraction between electrons in polar crystals can overcome
the Coulomb repulsion only when taking into account a \emph{dynamic}
electron-phonon response through the frequency-dependent dielectric function
\cite{Kirzhnits}. Moreover, the dynamic electron-electron interaction can
\emph{cooperate} with the electron-phonon pairing interaction
\cite{Akashi,Leggett1999}.

We consider the general case for the ratios of the LO-phonon energies to the
thermal and Fermi energies, without assuming adiabatic or antiadiabatic
limits. The calculation accounting for a multiband structure of the conduction
band is based on the gap equation from Ref. \cite{Takada1} for the gap
parameter $\Delta_{\lambda}\left(  \mathbf{k}\right)  $ (neglecting the
interband Josephson coupling, because it is not known and apparently does not
strongly influence the overall magnitude of $T_{c}$):%
\begin{align}
\Delta_{\lambda}\left(  \mathbf{p}\right)   &  =-\frac{1}{\left(  2\pi\right)
^{3}}\int d\mathbf{k}~\Delta_{\lambda}\left(  \mathbf{k}\right)  \frac
{\tanh\frac{\beta\left\vert \varepsilon_{\mathbf{k},\lambda}\right\vert }{2}%
}{2\left\vert \varepsilon_{\mathbf{k},\lambda}\right\vert }\nonumber\\
&  \times\left[  V_{\lambda}^{0}\left(  \mathbf{p}-\mathbf{k}\right)
+\frac{2}{\pi}\int_{0}^{\infty}d\Omega\frac{\operatorname{Im}V_{\lambda}%
^{R}\left(  \mathbf{p}-\mathbf{k},\Omega\right)  }{\Omega+\left\vert
\varepsilon_{\mathbf{k},\lambda}\right\vert +\left\vert \varepsilon
_{\mathbf{p},\lambda}\right\vert }\right]  , \label{gapeq1}%
\end{align}
where $\lambda$ is the index of the subband of the conduction band,
$\varepsilon_{\mathbf{k},\lambda}$ is the electron energy counted from the
chemical potential, $V_{\lambda}^{R}\left(  \mathbf{q},\Omega\right)  $ is the
effective electron-electron interaction potential, and $V_{\lambda}^{0}\left(
\mathbf{q}\right)  $ is its high-frequency limit. The effective interaction
potential takes into account both the Coulomb and retarded phonon-mediated
interactions, being expressed through the total dielectric function of the
electron-phonon system $\varepsilon_{\lambda}^{R}\left(  q,\Omega\right)  $:%
\begin{equation}
V_{\lambda}^{R}\left(  \mathbf{q},\Omega\right)  =\frac{4\pi e^{2}}%
{q^{2}\varepsilon_{\lambda}^{R}\left(  \mathbf{q},\Omega\right)  }. \label{Vr}%
\end{equation}
Here, $\varepsilon_{\lambda}^{R}\left(  \mathbf{q},\Omega\right)  $ is the
total dielectric function of the electron-phonon system for the $\lambda$-th
band. We use the dielectric function accounting for multiple LO-phonon
branches as treated in Ref. \cite{KDM2010}. The total dielectric function
(including both the lattice and electron polarization) is calculated within
the random phase approximation (RPA).

The trial gap parameter is modeled by the function $\Delta_{\lambda}\left(
\omega\right)  $ depending on the energy. The integral over the electron
momentum $\mathbf{k}$ can be transformed to the integral with the density of
states in the $\lambda$-th band $\nu_{\lambda}\left(  \varepsilon\right)  $:%
\begin{equation}
\frac{1}{\left(  2\pi\right)  ^{3}}\int d\mathbf{k}\ldots=\frac{1}{4\pi}\int
d\varepsilon~\nu_{\lambda}\left(  \varepsilon\right)  \int do_{\varepsilon
}\ldots
\end{equation}
and hence we arrive at the gap equation%
\begin{equation}
\Delta_{\lambda}\left(  \omega\right)  =-\int_{-\varepsilon_{F}}^{\infty}%
\frac{d\omega^{\prime}}{2\omega^{\prime}}\tanh\left(  \frac{\beta
\omega^{\prime}}{2}\right)  K_{\lambda}\left(  \omega,\omega^{\prime}\right)
\Delta_{\lambda}\left(  \omega^{\prime}\right)  \label{gapeq}%
\end{equation}
with the kernel%
\begin{align}
K_{\lambda}\left(  \omega,\omega^{\prime}\right)   &  =\frac{\nu_{\lambda
}\left(  \omega^{\prime}\right)  }{\left(  4\pi\right)  ^{2}}\int do_{\omega
}\int do_{\omega^{\prime}}\left[  V_{\lambda}^{0}\left(  \mathbf{p}_{\omega
}-\mathbf{p}_{\omega^{\prime}}\right)
\genfrac{}{}{0pt}{0}{{}}{{}}%
\right. \nonumber\\
&  \left.  +\frac{2}{\pi}\int_{0}^{\infty}d\Omega\frac{\operatorname{Im}%
V_{\lambda}^{R}\left(  \mathbf{p}_{\omega}-\mathbf{p}_{\omega^{\prime}}%
,\Omega\right)  }{\Omega+\left\vert \omega\right\vert +\left\vert
\omega^{\prime}\right\vert }\right]  . \label{gapeq2}%
\end{align}
Here, the moments $\mathbf{p}_{\omega}$ for a given energy $\omega$ are
restricted by the equation for an isoenergetic surface $\varepsilon
_{\mathbf{p},\lambda}=\omega$.

The gap equation (\ref{gapeq}) provides the energy-dependent gap parameter,
accounting then for a dynamic response of the electron-phonon system. It
appears that the kernel function is always positive \cite{DKTM2014}. However,
a significant frequency dependence of $K_{\lambda}\left(  \omega
,\omega^{\prime}\right)  $ does allow for non-trivial solutions of the
superconducting gap equation. These solutions come along with a significant
variation of the order parameter $\Delta_{\lambda}\left(  \omega\right)  $ as
a function of frequency, where a sign change of $\Delta_{\lambda}\left(
\omega\right)  $ can appear near the Fermi energy. This behavior of the gap
parameter might be visible in tunneling experiments.

\section{Discussion and summary}

The novelty of the present study with respect to the preceding works consists
in using the multimode dielectric function for SrTiO$_{3}$ with experimentally
determined LO and TO frequencies and in using the multiband model for the
electron conduction band with reliable band parameters corresponding to recent
experimental \cite{Meevasana} and theoretical \cite{VDM2011} works on the band
structure of SrTiO$_{3}$. For the numeric calculation, the high-frequency
dielectric constant $\varepsilon_{\infty}\approx5.44$ and the LO and TO phonon
frequencies in SrTiO$_{3}$ that we use are the same as those listed in our
paper \cite{KDM2010}, taken from literature sources (no fitting). The only
parameter which is not yet known definitely is the absolute deformation
potential $D$ for the electron -- acoustic-phonon interaction. Here, as in our
paper on superconductivity in a LAO-STO structure, we compare results for
three physically reasonable values:\ $D=3%
\operatorname{eV}%
$, $D=4%
\operatorname{eV}%
$ and $D=5%
\operatorname{eV}%
$. The value $D=4%
\operatorname{eV}%
$ was reported in Ref. \cite{Janotti}. The value $D=3%
\operatorname{eV}%
$ is very close to that (2.87 eV) found in Ref. \cite{Morozovska}.

The electrons in the conduction band are described by the matrix Hamiltonian
from Ref. \cite{VDM2011}
\begin{equation}
H_{ij}=\delta_{i,j}\varepsilon_{j}\left(  \mathbf{k}\right)  +W_{ij}/2
\end{equation}
with the energies%
\begin{equation}
\varepsilon_{j}\left(  \mathbf{k}\right)  =4t_{\pi}\sum_{i\neq j}\sin
^{2}\left(  \frac{a_{0}k_{i}}{2}\right)  +4t_{\delta}\sin^{2}\left(
\frac{a_{0}k_{j}}{2}\right)  ,
\end{equation}
where $a_{0}$ is the lattice constant. The matrix $W$
\begin{equation}
W=\left(
\begin{array}
[c]{ccc}%
2d & \xi & \xi\\
\xi & 2d & \xi\\
\xi & \xi & -4d
\end{array}
\right)
\end{equation}
describes the mixing of subbands within the conductivity band. We use the
values of the band parameters $t_{\delta},t_{\pi,}d,\xi$ from Ref.
\cite{VDM2011}: $t_{\delta}=35$ meV, $t_{\pi}=615$ meV, $\xi=18.8$ meV and
$d=2.2$ meV.

Next, we take into account the fact that even for the highest concentration of
electrons at which the superconductivity in $n$-doped SrTiO$_{3}$ is observed
(about $n_{0}\approx4\times10^{20}$ cm$^{-3}$), the Fermi energy of electrons
(counted from the bottom of the lowest band) is $E_{F}\sim0.1$ eV, which is
very small with respect to the half-width of the conduction band ($\sim2.5$
eV). Therefore we can use the parabolic approximation for each subband
determining the parameters from the tight-binding model \cite{VDM2011}
described above. Following Ref. \cite{Takada1}, the band mass used in the
present calculation is the density-of-state band mass $m_{D}=\left(
m_{xx}m_{yy}m_{zz}\right)  ^{1/3}$. The values of $m_{D,\lambda}$ appear to be
different for different $\lambda$: $m_{D,1}\approx0.669m_{e}$, $m_{D,2}%
\approx0.622m_{e}$, and $m_{D,3}\approx0.595m_{e}$, where $m_{e}$ is the
electron mass in vacuum. The inverse-averaged band mass [determined as
$m_{b}^{-1}=\left(  m_{xx}^{-1}+m_{yy}^{-1}+m_{zz}^{-1}\right)  /3$] is with
high accuracy one and the same for three bands: $m_{b}\approx0.593m_{e}$.

\begin{figure}
[th]
\begin{center}
\includegraphics[
height=2.303in,
width=3.1886in
]%
{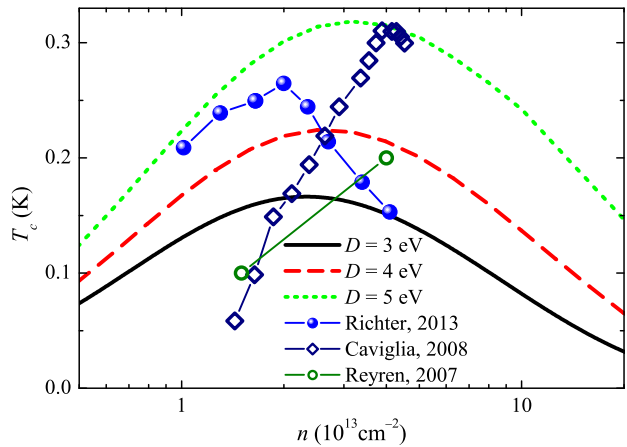}%
\caption{Critical temperature for the superconducting phase transition for the
electron gas on the SrTiO$_{3}$/LaAlO$_{3}$ interface, compared to the
experimental data of Refs. \cite{Reyren,Caviglia,Richter}.}%
\label{laosto}%
\end{center}
\end{figure}

The critical temperatures for the superconducting phase transition on the
SrTiO$_{3}$/LaAlO$_{3}$ interface have been calculated in Ref. \cite{DKTM2014}%
. In Fig. \ref{laosto}, we show the the critical temperatures vs the
two-dimensional electron concentration. The results are compared to the
experimental data of Refs. \cite{Reyren,Caviglia,Richter}.%

\begin{figure}
[th]
\begin{center}
\includegraphics[
height=2.511in,
width=3.2019in
]%
{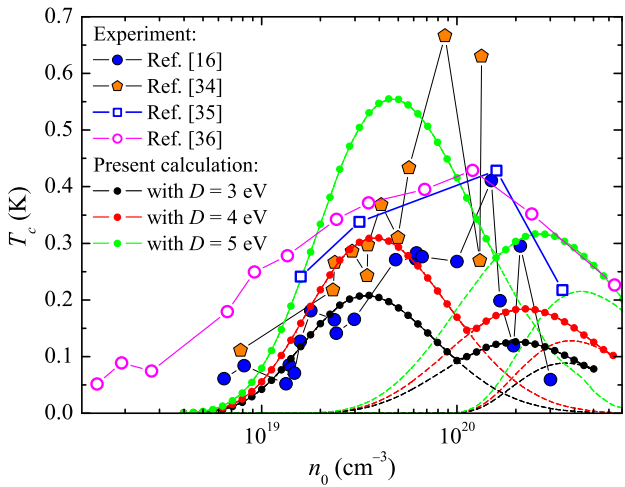}%
\caption{Critical temperature for the superconducting phase transition in
$n$-doped SrTiO$_{3}$ as a function of the carrier density. The results of the
present calculation for three values of the deformation potential $D$ are
shown by full dots. The calculated critical temperatures are compared with the
experimental data \cite{Koonce,Binnig,Lin2,Schooley} shown by symbols.}%
\label{stobulk}%
\end{center}
\end{figure}

The numeric results for the critical temperatures in bulk strontium titanate
as a function of the electron concentration are shown in Fig. \ref{stobulk}.
When neglecting the Josephson interband coupling, the actual critical
temperature is determined as the highest $T_{c}$ of the solutions of the gap
equation (\ref{gapeq}) for three subbands of the conduction band. These
particular solutions are shown in the figure by dashed curves, and the
resulting $T_{c}$ for each $D$ are shown by the curves with full dots.

Qualitatively, the measured critical temperatures in different experiments,
both for bulk strontium titanate and for its interface, lie in the same range
of the carrier concentrations and have the same range of magnitude. However,
there exists a significant discrepancy between different experimental data
\cite{Koonce,Binnig,Lin2,Schooley}. These discrepancies of the experimental
results has a transparent explanation. The thermal energy is very small with
respect to the Fermi energy of the electrons and the LO-phonon energies.
Therefore the critical temperatures can be very sensitive to relatively small
difference of the material parameters of the experimental samples. Thus even a
relatively small change of these parameters can then lead to a significant
change of the critical temperature. Moreover, experimental data are obtained
with some numeric inaccuracy which also influences the results.%

A small dome at lower density observed in Ref. \cite{Schooley} is not captured
by the present calculation due to a simplified model of the band structure.
However, the obtained density dependence of the critical temperature in
$n$-doped SrTiO$_{3}$ is in line, at least qualitatively, with the existing
experimental data without fitting. Moreover, the calculated critical
temperature, as well as the experimental results, correspond to the physical
picture of multiband superconductivity in strontium titanate.

In summary, the dielectric function method provides a reliable explanation of
superconductivity in strontium titanate, both in bulk and at its interface. In
the dielectric response of the electron-phonon system, all phonon branches are
taken into account. The interplay between the dynamic screening of the Coulomb
interaction, as described by the pairing kernel function $K_{\lambda}\left(
\omega,\omega^{\prime}\right)  $ and the values of the Fermi energy
$\varepsilon_{F}$ can provide a non-monotonic dome-shape of the critical
temperature both in bulk SrTiO$_{3}$ and on the SrTiO$_{3}$/LaAlO$_{3}$
interface, being in line with experimental data. Despite the apparent
uncertainty of the experimental results on the critical temperatures, the
theoretical treatment of the superconducting phase transition in bulk
strontium titanate and on the SrTiO$_{3}$/LaAlO$_{3}$ interface leads to
qualitative agreement with experiment without fitting material parameters. The
calculated critical temperatures, as well as the experimental results, agree
with the physical picture of multiband superconductivity, and support the
hypothesis that the mechanism of superconductivity in SrTiO$_{3}$ and on the
SrTiO$_{3}$/LaAlO$_{3}$ interface is provided by the electron - optical-phonon interaction.

\begin{acknowledgements}
We are grateful to J. Ruhman for valuable discussions.
This work is supported by the Flemish Research Foundation (FWO-Vl), project
nrs. G.0115.12N, G.0119.12N, G.0122.12N, G.0429.15N, by the Scientific
Research Network of the Research Foundation-Flanders, WO.033.09N, and by the
Research Fund of the University of Antwerp.
\end{acknowledgements}

\end{document}